\documentclass[twocolumn,showpacs,preprintnumbers,prb,fleqn,floatfix,superscriptaddress]{revtex4-1}

\usepackage{amssymb}
\usepackage{epsfig}
\usepackage{amsmath}
\usepackage[breaklinks]{hyperref}%hyperef package does internal and external linking (labels, urls)
\usepackage[all]{hypcap}%addition to hyperref, it ensures figure is correctly displayed (and not only the caption) when link is clicked

\hypersetup{
    plainpages=false,
    bookmarks=false,         % show bookmarks bar?
    unicode=false,          % non-Latin characters in Acrobat's bookmarks
    pdfborder={0 0 0}
    pdftoolbar=true,        % show Acrobat's toolbar?
    pdfmenubar=true,        % show Acrobat's menu?
    pdffitwindow=false,     % window fit to page when opened
    pdfstartview={FitH},    % fits the width of the page to the window
    pdfnewwindow=true,      % links in new window
    linktoc=section,
    colorlinks=true,       % false: boxed links; true: colored links
    linkcolor=blue,          % color of internal links
    citecolor=blue,        % color of links to bibliography
    filecolor=magenta,      % color of file links
    urlcolor=blue           % color of external links
}

\begin{document}

\title{High field magnetospectroscopy to probe the $1.4$~eV Ni color center in diamond}

\author{P. \surname{Plochocka}}
\email{paulina.plochocka@lncmi.cnrs.fr} \affiliation{Laboratoire National des Champs Magn\'etiques Intenses,
CNRS-UJF-UPS-INSA, 143, avenue de Rangueil, 31400 Toulouse}\affiliation{Laboratoire National des Champs Magn\'etiques
Intenses, CNRS-UJF-UPS-INSA, 25 rue des Martyrs 38042 Grenoble, France}

\author{O. \surname{Portugall}}
 \affiliation{Laboratoire National des Champs Magn\'etiques Intenses,
CNRS-UJF-UPS-INSA, 143, avenue de Rangueil, 31400 Toulouse}

\author{P. Y.  \surname{Solane}}
 \affiliation{Laboratoire National des Champs Magn\'etiques Intenses,
CNRS-UJF-UPS-INSA, 143, avenue de Rangueil, 31400 Toulouse}

\author{E.  \surname{Gheeraert}}
 \affiliation{Institut N\'eel, CNRS and Universit\'e Joseph Fourier, 25 rue des Martyrs 38042 Grenoble, France}
 \author{L.  \surname{Ranno}}
 \affiliation{Institut N\'eel, CNRS and Universit\'e Joseph Fourier, 25 rue des Martyrs 38042 Grenoble, France}
\author{E.  \surname{Bustarret}}
 \affiliation{Institut N\'eel, CNRS and Universit\'e Joseph Fourier, 25 rue des Martyrs 38042 Grenoble, France}
\author{N.  \surname{Bruyant}}
 \affiliation{Laboratoire National des Champs Magn\'etiques Intenses,
CNRS-UJF-UPS-INSA, 143, avenue de Rangueil, 31400 Toulouse}
\author{I.  \surname{Breslavetz}} \affiliation{Laboratoire
National des Champs Magn\'etiques Intenses, CNRS-UJF-UPS-INSA, 25 rue des Martyrs 38042 Grenoble, France}
\author{D. K.  \surname{Maude}} \affiliation{Laboratoire
National des Champs Magn\'etiques Intenses, CNRS-UJF-UPS-INSA, 25 rue des Martyrs 38042 Grenoble, France}

\author{H.  \surname{Kanda}}
\affiliation{National Institute for Materials Science, 1-1 Namiki,
305-0044 Tsukuba, Japan}
\author{G. L. J. A. \surname{Rikken}}
 \affiliation{Laboratoire National des Champs Magn\'etiques Intenses,
CNRS-UJF-UPS-INSA, 143, avenue de Rangueil, 31400 Toulouse}

\
\date{\today }
\pacs{71.55.-i, 78.55.-m}

\begin{abstract}
A magneto-optical study of the $1.4$~eV Ni color center in boron-free synthetic diamond, grown at high pressure and
high temperature, has been performed in magnetic fields up to $56$~T. The data is interpreted using the effective spin
Hamiltonian of Nazar\'e, Nevers and Davies [Phys. Rev. B \textbf{43}, 14196 (1991)] for interstitial Ni$^{+}$ with the
electronic configuration $3d^{9}$ and effective spin $S=1/2$. Our results unequivocally demonstrate the trigonal
symmetry of the defect which preferentially aligns along the [111] growth direction on the (111) face, but reveal the
shortcomings of the crystal field model for this particular defect.
\end{abstract}

\maketitle

\section{Introduction}

Diamond, as a material, has attracted a lot of attention due to its unique physical properties; it is the hardest known
material with high thermal conductivity and a $5.5$~eV wide electronic band gap. The large Debye temperature of diamond
reduces the interaction between impurities and the lattice leading to almost atomic like optical emission spectra of
defects with extremely narrow lines. The so-called color centers, various transition metal-nitrogen/vacancy complexes
in diamond, can act as single photon sources, capable of photostable operation at room
temperature,~\cite{Aharonovich09,Aharonovich09a,Simpson09,Castelletto10} with possible applications in quantum
information processing~\cite{Morton06}. Moreover, the nitrogen/vacancy (NV) center in diamond has been used to image a
single electronic spin using nanoscale magnetometry.~\cite{Maze08, Balasubramanian08} The NV center has also been used
to produce diamond based light emitting diodes.~\cite{Lohrmann11} These developments have led to a renewed interest in
the optical properties of color centers in diamond.

Macroscopic synthetic diamond crystals are mainly grown by chemical-vapor deposition (CVD) methods or by the
high-pressure high-temperature (HPHT) method, the latter giving mm$^3$ size bulk crystals. In both growth techniques
there are only a few impurities which can enter into the diamond structure. Using the HPHT method, the incorporation of
cobalt (Co)~\cite{Lawson96} or nickel (Ni)~\cite{Collins82, Zaitsev00} has been achieved in significant amounts.
Incorporating transition metals into diamond is of interest for applications such as single photon emitters~\cite{
Aharonovich09, Aharonovich09a, Simpson09, Castelletto10} or spintronics in analogy to diluted magnetic semiconductors
(DMS). The Curie temperature of DMS has been predicted to scale with the inverse cube of the lattice constant of the
host matrix~\cite{Dietl01}. Diamond has the smallest lattice constant $(a=3.56{\AA})$ of all semiconductors, making it
an excellent candidate for ferromagnetic ordering above room temperature. For all these reasons, a thorough
understanding of transition metal complexes in diamond has become essential to further develop diamond related
technologies.

Despite the numerous potential applications of the nickel color center in diamond, its exact crystallographic site and
electronic properties are still under debate. The NIRIM-2 electron spin resonance (ESR) line,\cite{Isoya90} has been
identified with the $1.4$~eV doublet of zero phonon lines (ZPLs) seen in optical
studies,\cite{Davies89,Nazare91,Iakoubovskii04,Mason99,Maes04} and unambiguously attributed to a nickel containing
center.\cite{Davies89,Nazare91} The ESR results suggested that the nickel is incorporated interstitially in a single
positively charge state $3d^{9}$, effective spin $S=1/2$ with trigonal symmetry and a strong trigonal distortion due to
the presence of an additional impurity or vacancy near by~\cite{Isoya90}. The trigonal symmetry was confirmed by
optical studies~\cite{Nazare91} under uniaxial stress and magnetic fields up to $6$~T. The agreement between the ESR
and magneto-optical data has lead some authors to propose that the Zeeman splitting of the $1.4$~eV line be used as a
pulsed magnetic field calibration probe.~\cite{Maes04} However, recent studies suggested an alternative complex of
nickel with boron,~\cite{Baker03} or even isolated interstitial Ni.~\cite{Larico04} Recent theoretical work also
contradicts the hypothesis of either isolated interstitial Ni or interstitial Ni complex with either a vacancy or an
impurity; the calculations predict that interstitial Ni with trigonal symmetry is unstable.~\cite{Larico09}

In order to further elucidate the nature of this defect, we have performed a magneto-optical study of the Ni color
center in synthetic diamond which is characterized by two zero-phonon lines (ZPLs) at $1.401$ and $1.404$~eV.
Measurements in high magnetic fields unequivocally demonstrate the trigonal symmetry of the defect which is mostly
incorporated on the (111) faces with its trigonal axis preferentially aligned along the growth direction. For
incorporation on the (001) face there is no preferential alignment of the defect axis along one of the four $<111>$
directions. These results are extremely well described by the effective spin Hamiltonian for a trigonal defect proposed
in the seminal paper of Nazar\'e, Nevers and Davies~\cite{Nazare91} (hereafter referred to as NND) using a single
parameter set for both the (111) and (001) faces. However, in certain regions of the (111) face the trigonal axis of
the defect is not preferentially aligned. Under such conditions, the correct parameters of the effective spin
Hamiltonian are significantly different.

\section{Sample characterization}

Two diamond crystals containing nickel and nitrogen have been studied. The crystals were grown in nickel solvent in
similar high pressure and high temperature conditions, except that for one of the crystals a nitrogen getter (Ti) was
added to the solvent in order to reduce the nitrogen incorporation. Crystal KA1970 (Sample A), grown with the nitrogen
getter has a light green color, and, KA1153 (Sample B) grown without the nitrogen getter has a yellow/brown color. As
there was no specific boron contamination of the growth environment, the concentration of this impurity is expected to
be negligible in the grown crystals. Both single crystals have mainly wide (111) growth sectors, with smaller (001)
growth sectors terminated by square faces. The type Ib diamond seed crystals were not removed. According to the growth
conditions, the nickel concentration is expected to be around $10^{19}$ cm$^{-3}$ in both crystals, and the nitrogen
concentration around $10^{17}$ cm$^{-3}$ in Sample A and $10^{19}$ cm$^{-3}$ in Sample B. Due to the dependence of the
impurity incorporation rate on the crystallographic orientation during growth, the impurity concentration is expected
to depend on the growth sector. Crystals were characterized by cathodoluminescence at $T=5$~K with a $30$~kV e-beam and
magnetometry using a SQUID magnetometer in the temperature range $2$ to $300$~K and in a $0$ to $5$~T magnetic field.

\subsection{Cathodoluminescence}

\begin{figure}
\includegraphics[width=85mm]{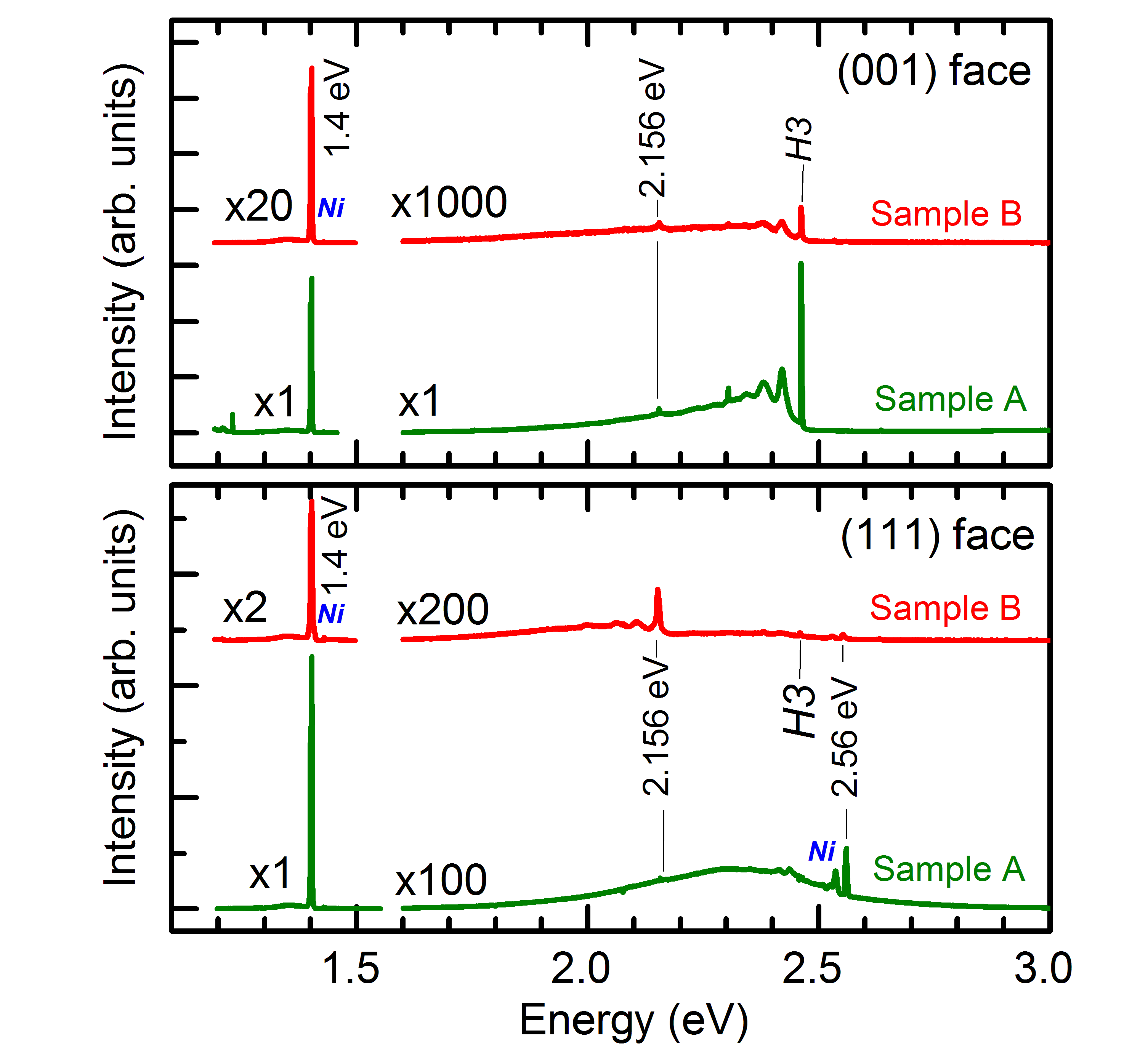}
\caption{\label{fig1_cathodo} (Color on-line) Cathodoluminescence spectra of Sample A (grown with N getter) and Sample
B (grown without N getter) recorded on a (100) and (111) growth sectors under similar conditions.}
\end{figure}

For Sample B, grown without the nitrogen getter, the cathodoluminescence spectra recorded on (100) and (111) growth
sectors exhibits mainly the $1.40$~eV center (Fig.~\ref{fig1_cathodo}), which can be attributed unambiguously to a
defect containing a single nickel atom~\cite{Davies89}. Nitrogen related centers H3 ($2.46$~eV, attributed to nitrogen
VNV complex) and $2.156$~eV (attributed to neutral NV complex) also appeared, but with stronger H3 signal on the (100)
sectors and stronger $2.156$~eV signal on $(111)$ sectors. A nickel related peak at $2.56$~eV was also observed on the
(111) growth sector, suggesting a stronger incorporation of Ni in the (111) growth sectors, in agreement with previous
studies. Note that no signal corresponding to W8 centers, related to substitutional Ni$_s^-$ were
observed.\cite{Isoya90a,Nazare01} For Sample A, grown with the nitrogen getter, the intensity of the $1.40$~eV lines is
significantly larger compared to Sample B. Spectra recorded on the $(111)$ growth sector reveal the $2.56$~eV Ni
related center, with very weak N related signals. This suggests a much stronger incorporation of Ni in $(111)$ growth
sectors compared to $(100)$.

\subsection{Magnetization}

\begin{figure}[t]
\includegraphics[width=85mm]{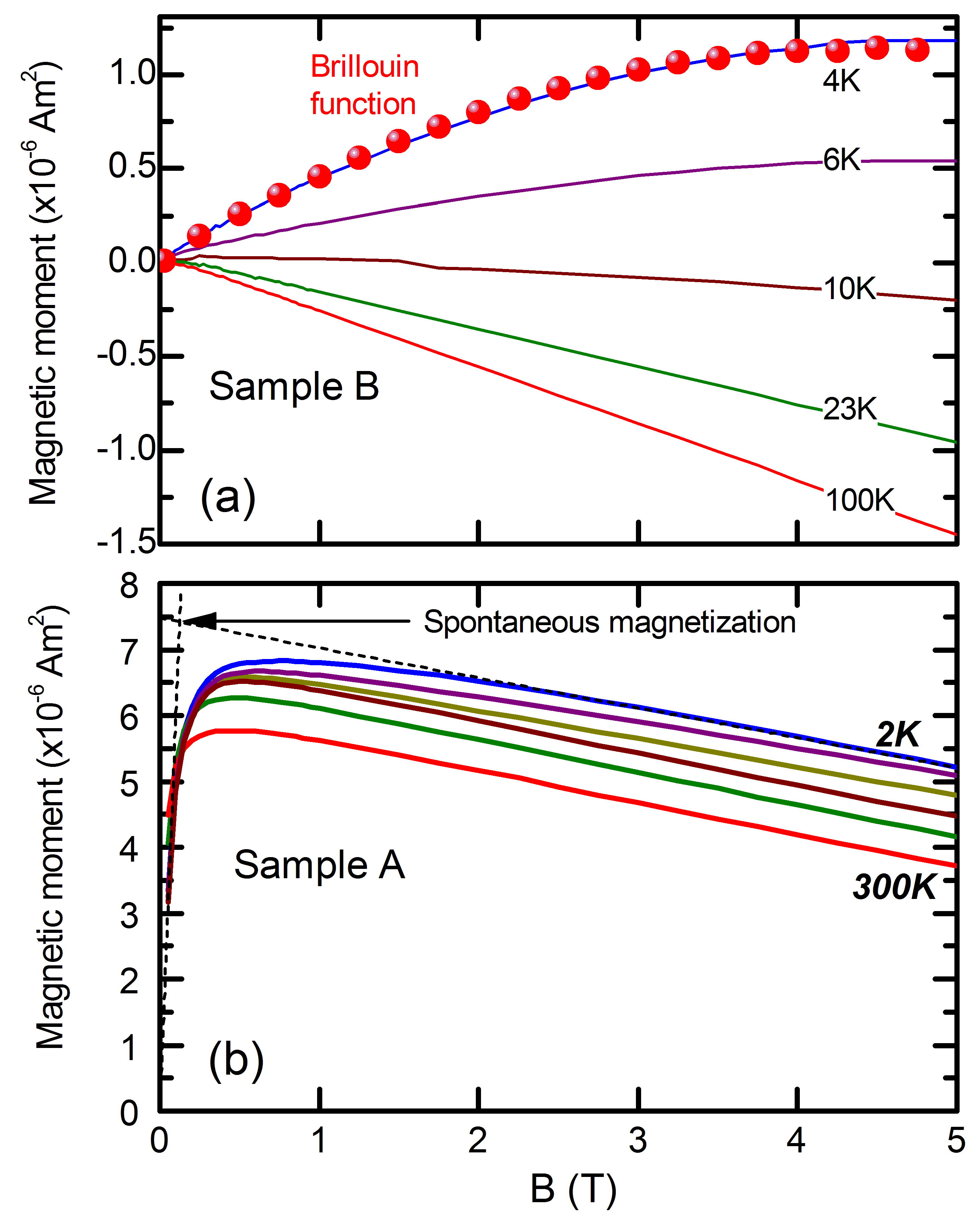}
\caption{\label{fig2_squid} (Color on-line) (a) SQUID measurement of Sample B. The symbols show a fit using a Brillouin
function as described in the text. (b) SQUID measurement of Sample A at temperatures $300, 200, 50, 10, 4, 2$~K. The
broken lines indicates method used to determine the spontaneous magnetization $M(T)$.}
\end{figure}

Information concerning the incorporation of Ni can also be obtained from magnetization measurements. Between room
temperature and $\simeq 100$~K, the magnetic moment of Sample B is proportional to field (Fig.~\ref{fig2_squid}(a))
with a negative slope, \emph{i.e.} negative susceptibility of $\chi=-2.5\times10^{-5}$. This is close to the
diamagnetic susceptibility of diamond ($\chi=-2.2\times10^{-5}$) and as expected shows no temperature dependence. Below
$100$~K an additional contribution appears, increasing in intensity as the temperature decreases. This contribution is
attributed to a paramagnetic component. This assignment is confirmed by the good agreement (see
Fig.\ref{fig2_squid}(a)) with a Brillouin function approach,
\begin{equation}
B_J(x)=\frac{2J+1}{2J}\coth\left(\frac{2J+1}{2J}x\right)-\frac{1}{2J}\coth\left(\frac{x}{2J}\right),\\
\end{equation}
with
\begin{equation}
x=\frac{gJ\mu_B B}{kT}.
\end{equation}
The total magnetic moment being
\begin{equation}
m(B)=N_{Ni}gJ\mu_B B_J(x),
\end{equation}
where $J$ is the Ni quantum number, $g$ the Land\'e factor, $\mu_B$ the Bohr magneton, $k$ the Boltzmann constant, $T$
the temperature, $N_{Ni}$ the concentration of paramagnetic centers and $B$ the applied magnetic field. The fit was
performed assuming $gJ=2$ and leaving $N_{Ni}$ as the only free parameter. A value of $gJ=2$ corresponds to isolated
nickel in the Ni$_2^+$ oxidation state ($3d^8$), assuming $J\approx S\approx 1$ and $g=2$. A concentration of
paramagnetic centers $N_{Ni}=1.5\times10^{19}$ cm$^{-3}$ was deduced. Note, a value of $gJ=3$ ($J=3/2$) or lower value
$gJ=1$ ($J=1/2$) would reduce or increase $N_{Ni}$ accordingly. Nevertheless, this concentration is in the order of
magnitude of the expected incorporated Ni concentration. Such a value corresponds to a Ni relative concentration of
$8\times10^{-5}$ \emph{i.e.} a very diluted magnetic system in which weak interactions between magnetic ions and a
paramagnetic behavior are expected. We deduce that nickel is incorporated as isolated, non interacting paramagnetic
centers.

\begin{figure}[ht]
\includegraphics[width=85mm]{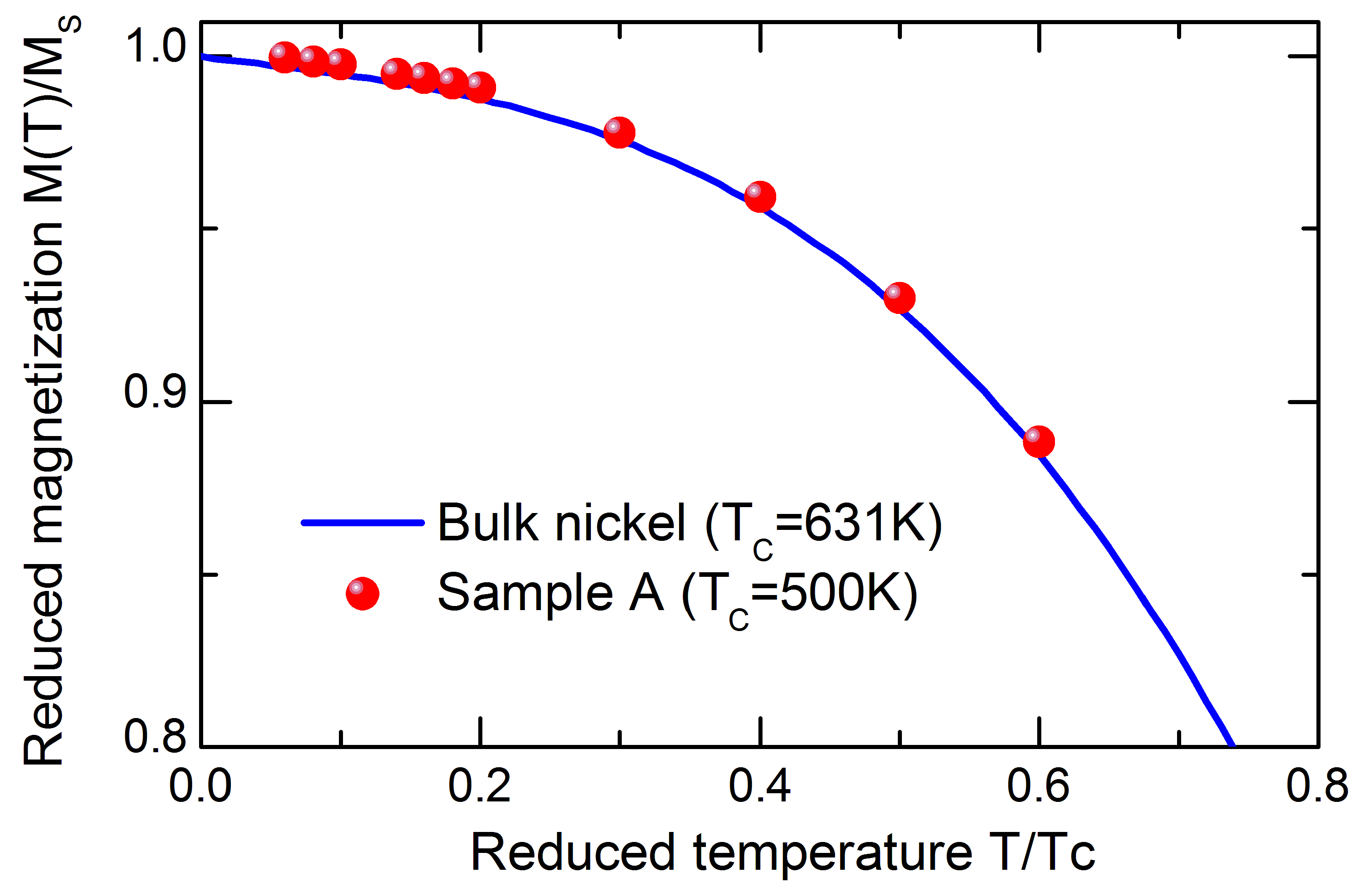}
\caption{\label{fig3_sponmag} (Color on-line) Reduced magnetic moment $M(T)/M_S$ versus reduced temperature for bulk
nickel (line) and Sample A (symbols), drawn for $T_c=631$ K for nickel and $T_c=500$ K for Sample A.}
\end{figure}

For Sample A the magnetic moment has a very different behavior, as seen in Fig.\ref{fig2_squid}(b). A magnetic field as
low as $0.5$~T is already sufficient to saturate the magnetic moment even at $300$~K, suggesting a ferromagnetic
behavior. The linear decrease observed for higher fields is attributed, as before, to the diamagnetic contribution of
the diamond matrix. Plots representing the reduced magnetization $M(T)/M_S$ as a function of the reduced temperature
$T/T_c$ should be similar for bulk nickel and for this ferromagnetic sample. Under such an assumption, the temperature
dependence of the experimental magnetization (Fig.~\ref{fig3_sponmag}) corresponds to a $500$~K Curie temperature, 20\%
smaller than the one of bulk nickel ($631$~K). This suggests the presence of nickel clusters, with a reduced Curie
temperature due to size-effects. The low temperature spontaneous magnetic moment of $M_S=7.4\times10^{-6}$ Am$^2$ gives
a nickel concentration of $N_{Ni}=3.5\times10^{19}$ cm$^{-3}$, assuming a $gJ=2$, which has the same order of magnitude
as for Sample B.

The strikingly different magnetic behavior of both samples indicates the strong influence of nitrogen on the
incorporation of Ni in diamond. Without nitrogen (\emph{i.e.} with a much lower N concentration than Ni concentration),
as in Sample A, Ni tends to form clusters, reducing the concentration of isolated nickel and the crystal is
ferromagnetic. With nitrogen, no ferromagnetic behavior is observed, the nickel atoms are diluted into the crystal as
paramagnetic centers. This suggests a lower formation energy for Ni-N complexes compared to Ni defects such as
interstitial Ni, substitutional Ni and NiV. This is in agreement with first principles theoretical investigation of
nickel related complexes in diamond\cite{Larico09}.

\section{Effective spin Hamiltonian}

\begin{figure}[b]
\begin{center}
\includegraphics[width=0.7\linewidth]{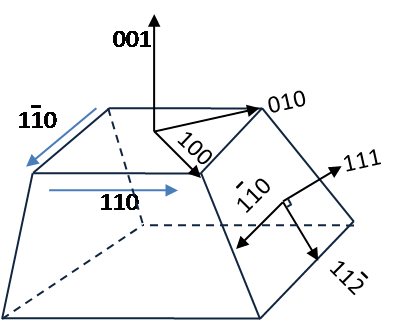}
\end{center}
\caption{(Color on-line) Simplified schematic of the HTHP synthetic diamond crystals labeling the principal
crystallographic axes. The small (113) and (101) faces of our cubo-octahedral crystal have been omitted for
clarity.}\label{fig4_diamond}
\end{figure}

Before presenting the magneto-photoluminescence results we briefly outline NND effective spin
Hamiltonian,~\cite{Nazare91} for trigonal interstitial Ni$^{+}$ with the electronic configuration $3d^{9}$ and
effective spin $S=1/2$, which is required to understand the data. The principle crystallographic orientations of the
HPHT diamond crystals investigated here are presented schematically in Fig~\ref{fig4_diamond}. It is possible to define
four sets of local axes $Z,X,Y$ (see Table~\ref{tab_localaxis}) corresponding to the four different possible
$\langle111\rangle$ orientations of the trigonal axis of the defect. The quantization axis $Z$ is parallel to the
$C_{3}$ axis. The perturbation of the magnetic field is given by the following Hamiltonian,
\begin{equation}\label{Hamiltonian}
    \Delta H=\mu_{B}(g_{1}B_{X}S_{X}+g_{1}B_{Y}S_{Y}+g_{3}B_{Z}S_{Z}),
\end{equation}
where $\mu_{B}$ is the Bohr magneton and the quantization axis is the $Z$ trigonal axis of the center.

The energy separation of the excited and ground state is large enough to ignore any interaction between them. The
secular matrix describing the perturbation of the magnetic field on the ground state doublet is,
\begin{widetext}

 \begin{equation}
  \label{matrix_ground}
   \frac{1}{2}\left(
     \begin{array}{cccc}
       g'_{3} \mu_B B_{Z} + \lambda & 0 & 0 & g_{1} \mu_B B_{X} - i g_{1} \mu_B B_{Y}\\
       0 & - g'_{3} \mu_B B_{Z} + \lambda & g_{1} \mu_B B_{X} - i g_{1} \mu_B B_{Y} & 0 \\
       0 & g_{1} \mu_B B_{X} + i g_{1} \mu_B B_{Y} & g_{3} \mu_B B_{Z} - \lambda & 0 \\
       g_{1} \mu_B B_{X} + i g_{1} \mu_B B_{Y} & 0 & 0 & -g_{3} \mu_B B_{Z} - \lambda \\
     \end{array}
   \right)
 \end{equation}
\end{widetext}
where $\lambda \simeq 2.8~$meV is the spin orbit splitting of the ground state and the various $g$ terms are the
effective Land\'e $g$-factors. The secular matrix describing the excited state in a magnetic field is given by,
\begin{equation}
  \label{matrix_exc}
   \frac{1}{2}\mu_B\left(
     \begin{array}{cc}
       g^{e}_{3}B_{Z} & g^{e}_{1}B_{X} - i g^{e}_{1}B_{Y}\\
       g^{e}_{1}B_{X} + i g^{e}_{1}B_{Y} & -g^{e}_{3}B_{Z}\\
     \end{array}
   \right)
 \end{equation}
The secular matrixes can easily be diagonalized either numerically or analytically. We have done both and verified that
the results are identical. The analytic expressions for the ground and excited states are,
%\begin{widetext}
\begin{multline}\label{Eq_eigenvalues}
E^{g}_{1,2} = \frac{1}{2}\mu_B\bigg{[}\mp \frac{1}{2}B_{Z}(g_{3} - g'_{3})\\
- \sqrt{g_{1}^{2}(B^{2}_{X}+B^{2}_{Y})+(\lambda \pm \frac{1}{2}B_{Z}(g'_{3} + g_{3}))^{2}}\bigg{]}\\
E^{g}_{3,4} = \frac{1}{2}\mu_B\bigg{[}\mp\frac{1}{2}B_{Z}(g_{3} - g'_{3})\\
+ \sqrt{g_{1}^{2}(B^{2}_{X}+B^{2}_{Y})+(\lambda \pm \frac{1}{2}B_{Z}(g'_{3} + g_{3}))^{2}}\bigg{]}\\
E^{e}_{1,2} = \pm \frac{1}{2}\mu_B\sqrt{(g_{1}^{e})^{2}(B^{2}_{X}+B^{2}_{Y})+ B_{Z}^{2}(g^{e}_{3})^{2}}\\
\end{multline}

%matlab expressions
%E3=0.5*( -0.5*BZ*(g3-gp3) + sqrt(g1^2 * (BX.^2 + BY.^2) + (lamda + 0.5*BZ*(gp3+g3)).^2  ));
%
%E4=0.5*( 0.5*BZ*(g3-gp3) + sqrt(g1^2 * (BX.^2 + BY.^2) + (lamda - 0.5*BZ*(gp3+g3)).^2  ));
%
%E1=0.5*( -0.5*BZ*(g3-gp3) - sqrt(g1^2 * (BX.^2 + BY.^2) + (lamda + 0.5*BZ*(gp3+g3)).^2  ));
%
%E2=0.5*( 0.5*BZ*(g3-gp3) - sqrt(g1^2 * (BX.^2 + BY.^2) + (lamda - 0.5*BZ*(gp3+g3)).^2  ));

The magnetic field lifts the degeneracy of the ground and excited states which split into four and two levels
respectively. The energy of the possible optical transitions is simply the energy difference between all levels in the
$^2$A excited state and all levels in the $^2$E ground state. Thus, taking into account the four possible orientations
of the defect axis (see Table~\ref{tab_localaxis}), we expect a maximum of $32$ lines in the spectrum depending upon
the orientation of the magnetic field. When the magnetic field is aligned along a symmetry axis the number of
transitions with different energies is greatly reduced.

\begin{table}[h]
\begin{center}
%\Huge
\begin{tabular}{c | c c c}
 Label&Z&X&Y\\ %%
\hline
$\alpha$ & $[111]$ & $[11\bar2]$ & $[1\bar10] $\\
$\beta$ & $[1\bar11]$ & $[1\bar1\bar2]$ & $[\bar1\bar10]$ \\
$\gamma$ & $[\bar1\bar11]$ & $[\bar1\bar1\bar2]$ & $[\bar110]$\\
$\delta$ & $[\bar111]$ & $[\bar11\bar2]$ & $[110]$\\
\end{tabular}
%\normalsize
\end{center}
\caption{Summary of the four possible orientations of the local ($ZXY$) axis of the trigonal defect along the $<111>$
directions. The different orientations are labeled $\alpha, \beta, \gamma, \delta$.}\label{tab_localaxis}
\end{table}

\section{Magneto-photoluminescence}

\begin{figure}[t]
\begin{center}
\includegraphics[width=1\linewidth]{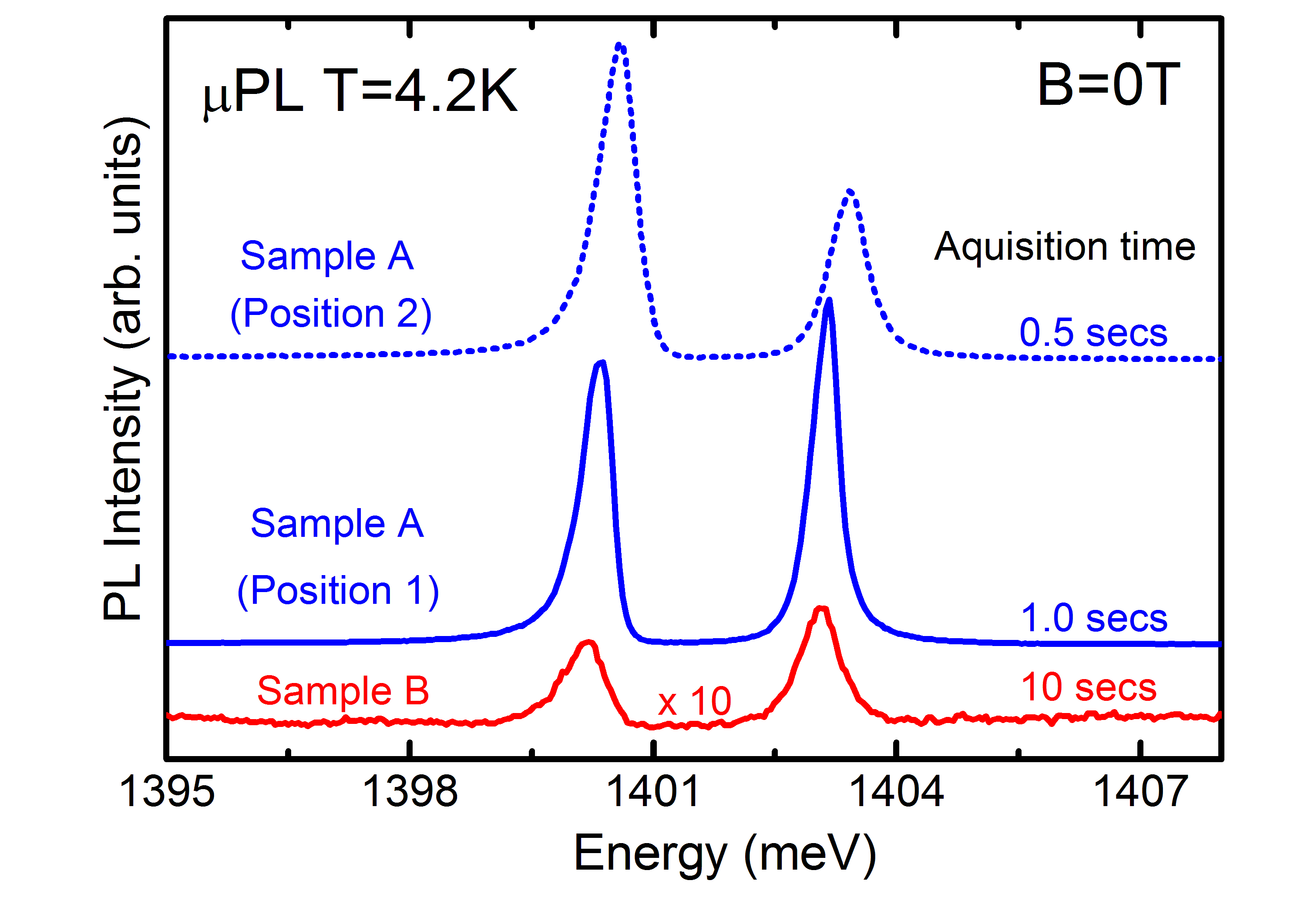}
\end{center}
\caption{(Color on-line) Typical $\mu$-photoluminescence spectra of Ni incorporated in HPHT synthetic diamond for
Samples A (at two different positions) and B at $T=4.2$~K and $B=0$~T. All spectra are taken on the (111) face. A
linear background has been subtracted from the rather weak PL signal of Sample B and the curves are offset vertically
for clarity. Position 2 is close to the bottom of the $(111)$ face, \emph{i.e.} near the seed crystal. See $\mu$-PL map
in Fig.~\ref{fig11_111map}.}\label{fig5_PL_0T}
\end{figure}

Micro photoluminescence ($\mu$PL) in \emph{dc} magnetic fields up to $28$~T and macro photoluminescence (PL) in pulsed
magnetic field up to $56$~T have been performed. For both experiments an optical fiber was used for the excitation and
collection of the emission from the sample.  A CW Ti:Sapphire laser tuned to $720$ nm or a CW solid state laser at
$660$~nm was used for the excitation. The emission spectra have been measured using a spectrometer equipped with a CCD
camera. For the macro PL the sample was placed at the end of the fiber with a diameter of 50 $\mu$m$^{2}$. In $\mu$PL
measurements light was focused on the sample using a microscope objective coupled with a mono-mode fiber. The size of
the laser spot on the sample was around $1\mu$m$^{2}$. The sample was mounted on piezo translation stages which allows
spectrally resolved spatial mapping in magnetic field. The measurements were performed at $T\simeq 4$~K with the
magnetic field applied parallel to (Faraday configuration) and normal to (Voigt configuration) the direction of
propagation of the light. A number of different orientations of the crystal with respect to magnetic field have been
measured on the (111) and (001) faces. Typical optical spectra, for both samples, measured at low temperature and at
zero magnetic field are presented in Fig~\ref{fig5_PL_0T}. A characteristic zero-phonon doublet is observed at $1.401$
and $1.404$ eV. The doublet structure originates from the $\lambda \simeq 2.8$~meV splitting of the $(^{2}E)$ ground
state due to a combination of spin orbit interaction and trigonal distortion. The asymmetric line shape is the result
of unresolved Ni isotopic splitting.~\cite{Nazare91} For sample A representative spectra recorded at two different
locations are shown. The small shift of the energetic position and the slightly different doublet splitting observed in
the spectra probably indicates indicates a different local strain.~\cite{Nazare91}

\begin{figure}[]
\begin{center}
\includegraphics[width=1\linewidth]{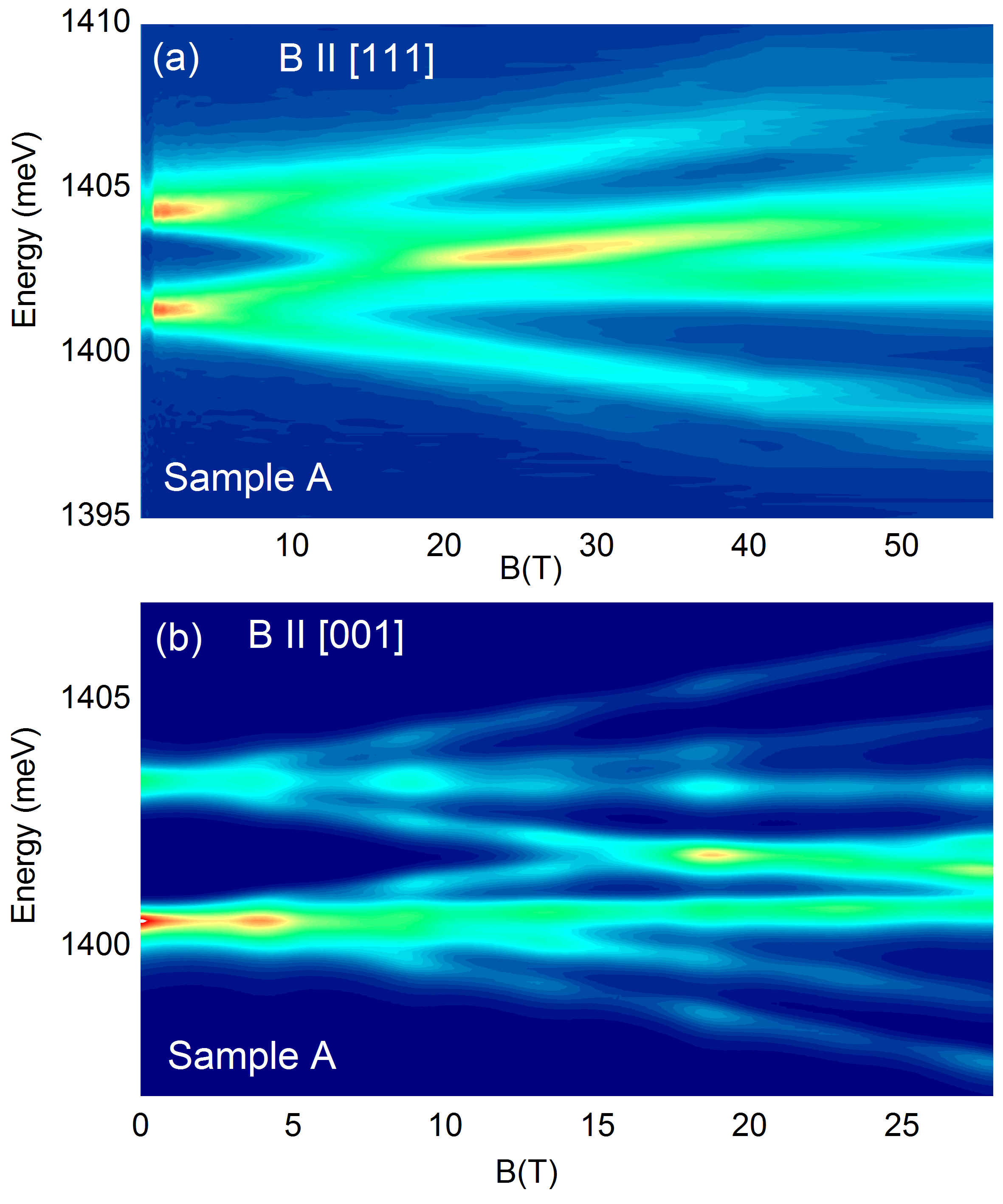}
\end{center}
\caption{(Color on-line) Representative $T=4.2$~K magneto-photoluminescence spectra in the Faraday configuration of the
$1.4$~eV Ni color center in HPHT synthetic diamond. (a) Macro-PL with $B \parallel [111]$ collected from the $(111)$
face. (b) $\mu$-PL with $B \parallel [001]$ collected from the $(001)$ face. Please note the different magnetic field
ranges.}\label{fig6_PLcolor}
\end{figure}

The intensity of the $1.4$~eV emission from sample B was too weak to be measured in pulsed magnetic field, where
typical integration time is of the order is $2$ms. Both samples were measured using $\mu$-PL in \emph{dc} magnetic
field up to $28$T. Selected representative macro and micro magneto-photoluminescence spectra for sample A are presented
in Fig~\ref{fig6_PLcolor}(a-b) respectively. The spectra were taken in the Faraday configuration on the (111) face with
$B
\parallel [111]$ (Fig~\ref{fig6_PLcolor}(a)) and on the (001) face with $B \parallel [001]$ (Fig~\ref{fig6_PLcolor} (b)). For
both experimental configuration we observe a splitting on the zero phonon doublet into multiple lines due to the Zeeman
effect. In order to compare the experimental results with the predictions of the effective spin Hamiltonian the energy
of each transition was extracted from the PL spectra by fitting a Gaussian function.

\subsection{Macro-photoluminescence on the (111) face}

\begin{figure}[t]
\begin{center}
\includegraphics[width=1\linewidth]{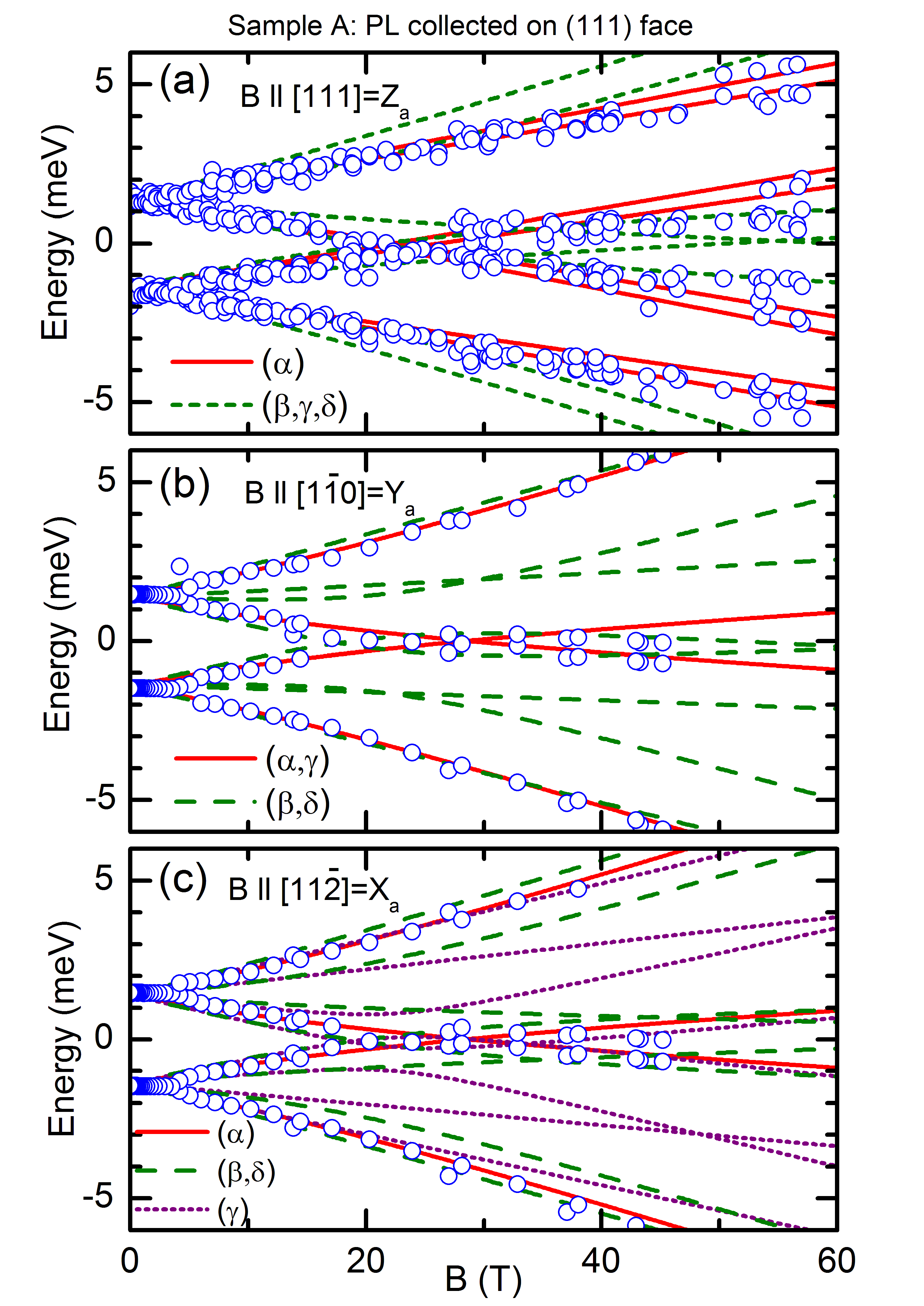}
\end{center}
\caption{(Color on-line) Energy of the transitions (symbols) for photoluminescence collected from the $(111)$ face of
the diamond crystal (Sample A) as a function of the magnetic field applied parallel to the $[111]$, $[1\bar{1}0]$ and
$[11\bar{2}]$ directions. The transition energies calculated with the effective spin Hamiltonian are shown as red solid
lines for a defect axis aligned along the $[111]$ growth direction of the face. The green broken lines are the
calculated energies for defects aligned along the other $\langle111\rangle$ directions. The legend labels the $Z$-axis
of the defect as in Table~\ref{tab_localaxis}.}\label{fig7_macro111}
\end{figure}

Using both Faraday and Voigt configurations as appropriate, the macro-photoluminescence has been collected on the (111)
face with the magnetic field aligned along the experimentally available symmetry axes of the defect (\emph{i.e.}
perpendicular to the face or parallel or perpendicular to certain edges of the face - see Fig.\ref{fig4_diamond}). The
energy of the observed optical transitions for $B\parallel[111]$, $B\parallel[1\bar10]$ and $B\parallel[11\bar2]$ is
plotted as symbols in Fig.\ref{fig7_macro111}(a-c). Only a few transitions are observed with most spectra showing only
four lines. The simplicity of the observed spectra immediately suggests that the defect axis is preferentially aligned
along the $[111]$ growth direction of the face. NND already reported a preferential alignment of the defect axis along
one of the $\langle111\rangle$ directions.

Making this assumption we fit the transitions energies calculated with the effective spin Hamiltonian. The
configuration $B\parallel[111]$ for a defect aligned along $[111]$, \emph{i.e.} the $Z$ axis, is a particularly easy
case to fit since all the off diagonal terms in the secular matrices are zero and only the parameters $g_3$ and
$g_3^{'} $ play a role ($g_3^e$ which controls the splitting of the excited state which is not resolved here is
determined later from the high resolution $\mu$-PL measurements). Using the NND parameters in
Ref.[\onlinecite{Nazare91}] as a starting point we have fitted the $B\parallel[111]$ data. Subsequently, the other
orientations, $B\parallel[1\bar10]$ and $B\parallel[11\bar2]$ were simulated to extract $g_1$ and $g_1^e$. The results
of such a fit are shown by the solid red lines in Fig.\ref{fig7_macro111}(a-c). The agreement is extremely good
apparently confirming the trigonal symmetry and preferential orientation of the defect axis along the $[111]$ growth
direction. For the $[111]$ aligned defect there is no difference between the $B\parallel[1\bar10]$ and
$B\parallel[11\bar2]$ orientations since these directions represent the local $Y$ and $X$ axes which are equivalent.
This is confirmed by the almost identical evolution of the observed transition energies for the two directions
(Fig.\ref{fig7_macro111}(b-c)).

For $B\parallel[1\bar10]$ the green broken lines are the calculated transition energies for defects with their axis
aligned along the $[1\bar11]$ and $[\bar111]$ directions. Clearly these transitions are not reproduced in the data.
However, from this orientation we can only say that the defect is either aligned along the $[111]$ or the
$[\bar1\bar11]$ directions for which the projection of the magnetic field onto the local axis have the same magnitude
so that the transitions are degenerate (red solid lines). It is the $B\parallel[11\bar2]$ data which tells us
definitively that the defect is aligned along $[111]$; the projection of the magnetic field is different for all the
other $\langle111\rangle$ directions and the calculated transitions for the $[\bar1\bar11]$ (purple dotted lines)
defect orientation are clearly not observed.

%Z axis 1=[1 1 1], 2=[-1 1 1], 3=[-1 -1 1], 4=[1 -1 1]

\subsection{Macro-photoluminescence on the (001) face}

\begin{figure}[t!]
\begin{center}
\includegraphics[width=1\linewidth]{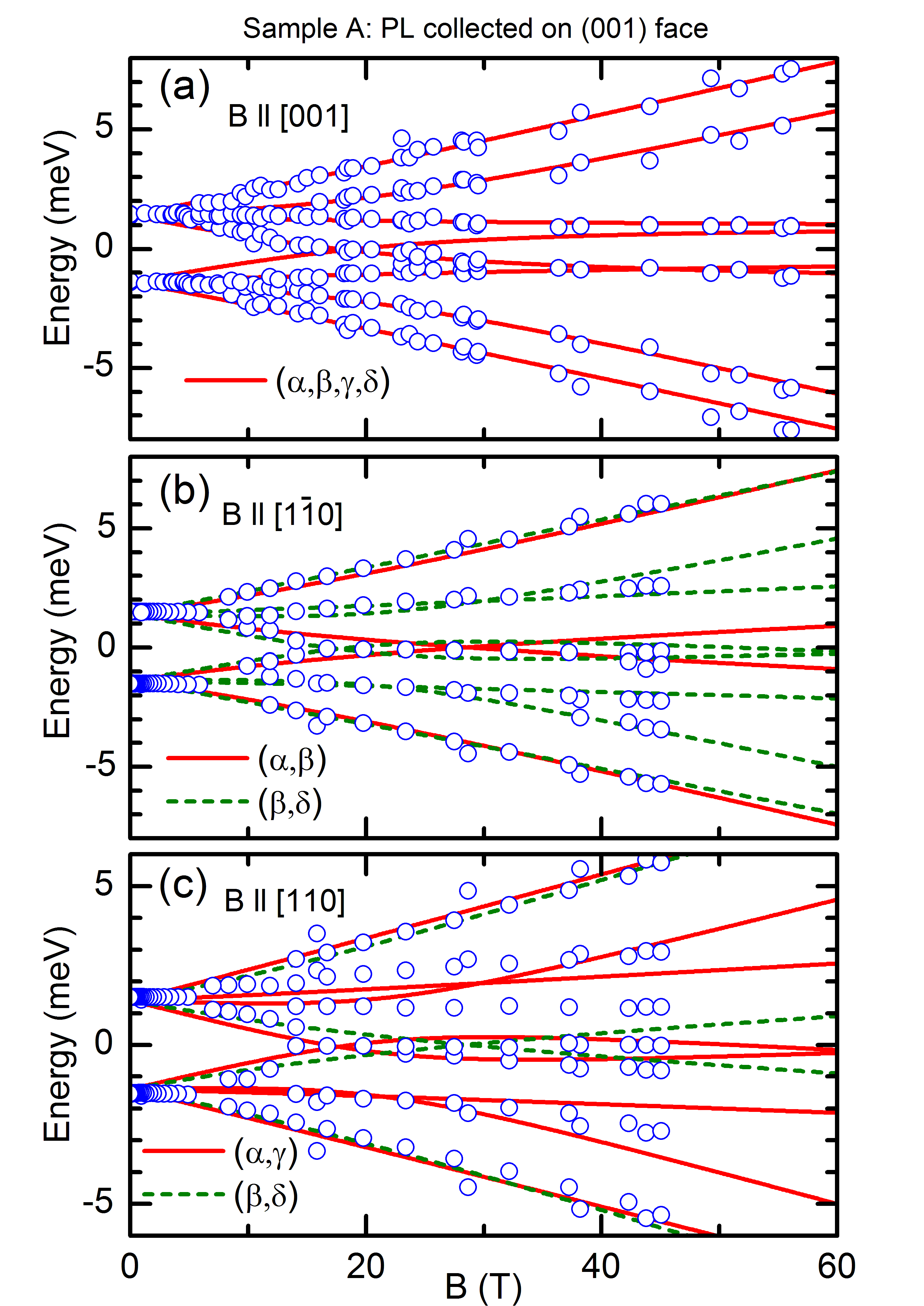}
\end{center}
\caption{(Color on-line) Energy of the transitions (symbols) for photoluminescence collected from the $(001)$ face of
the diamond crystal (Sample A) as a function of the magnetic field applied parallel to the $[001]$, $[1\bar{1}0]$ and
$[110]$ directions. The transition energies calculated with the effective spin Hamiltonian are shown as red solid lines
for a defect axis aligned along the $[111]$ growth direction of the face. The green broken lines are the calculated
energies for defects aligned along the other $\langle111\rangle$ directions. The legend labels the $Z$-axis of the
defect as in Table~\ref{tab_localaxis}.}\label{fig8_macro001}
\end{figure}
As before, using both Faraday and Voigt configurations, macro-photoluminescence has been collected on the $(001)$ face
with the magnetic field aligned along the experimentally available symmetry axes of the defect. The energy of the
observed optical transitions for $B\parallel[001]$, $B\parallel[1\bar10]$ and $B\parallel[110]$ is plotted as symbols
in Fig.\ref{fig8_macro001}(a-c). In contrast to the (111) face, we find that there is no preferential orientation of
the defect axis. The observed transitions have been simulated using the NND effective spin Hamiltonian. This fit was
performed simultaneously with the fit to the data on the (111) face, allowing a global optimization of the parameters.
The parameters found are summarized in Table~\ref{tab1}.

For the $B\parallel[001]$ all the possible defect alignments are equivalent so that no information concerning a
preferential orientation can be obtained form this data. Nevertheless, the predictions of the effective spin
Hamiltonian (solid red lines) fit the data very nicely. More information can be gained from the $B\parallel[1\bar10]$
and $B\parallel[110]$ orientations. In Fig.\ref{fig8_macro001}(b-c), the red solid lines correspond to defects aligned
along $[111]$ or $[\bar1\bar11]$ and the green dashed lines to defects oriented along $[\bar111]$ or $[1\bar11]$
directions. In the case of a preferential alignment one of the two $B\parallel[1\bar10]$ or $B\parallel[110]$
orientations would have an extremely simple spectrum composed of only four lines. This is clearly not the case; all
predicted transitions are observed for both directions. Thus, at least one of the $[111]$ or $[\bar1\bar11]$ directions
and at least one of the $[\bar111]$ or $[1\bar11]$ directions are occupied. The results on the (001) face suggest that
the preferential alignment along the [111] direction on the (111) face may be linked to the growth process.

\subsection{{$\mu$}-PL measurements}

The $\mu$-PL technique is not as useful for determining the symmetry of the defect since only the Faraday configuration
can be used so that the only orientation available is with the magnetic field perpendicular to the face under
investigation. It does however have certain advantages; (i) excitation and collection is very efficient which allows
the investigation of samples with a low emission intensity and (ii) spectrally resolved maps with a spatial resolution
$\simeq 1 \mu$m can be made given information concerning the homogeneity of the diamond crystal. The $\mu$-PL
measurements have been performed at $T=4.2$~K using a resistive magnet in static fields up to $28$~T.

\begin{figure}[t]
\begin{center}
\includegraphics[width=1\linewidth]{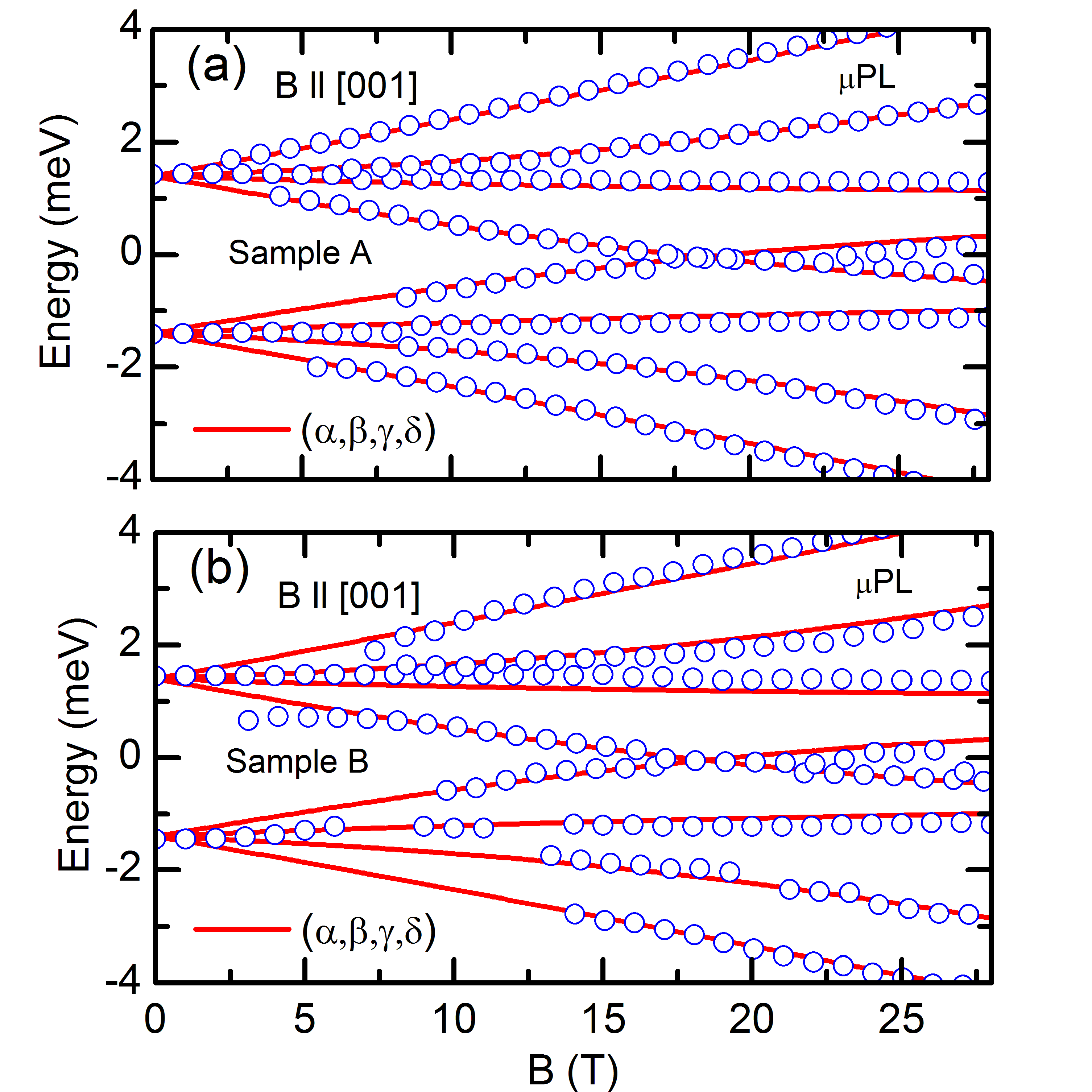}
\end{center}
\caption{(Color on-line) Energy of the observed transitions (symbols) for $T=4.2$~K $\mu$-photoluminescence collected
on the $(001)$ face with $B\parallel[001]$ on samples A and B as a function of the magnetic field. The red solid lines
are the predictions of the NND effective spin Hamiltonian (all defect alignments are equivalent for $B\parallel[001]$).
The legend labels the $Z$-axis of the defect according to Table~\ref{tab_localaxis}.}\label{fig9_micro001}
\end{figure}

In Fig.~\ref{fig9_micro001}(a-b) emission obtained from the (001) face of samples A and B is presented (symbols). The
results for the two different samples are almost identical despite the approximately two orders of magnitude larger
nitrogen concentration of sample B. As the emission intensity from sample B was very weak, fewer transitions are
resolved at low magnetic field. The solid red lines are the predictions of the effective spin Hamiltonian using exactly
the same parameters as before. As for the macro-PL measurements, the fits are very good for a data set of much higher
quality further confirming the trigonal symmetry of the defect. For the (001) face with $B\parallel[001]$ all the
defect orientations are equivalent so that no information concerning defect alignment can be extracted. The nickel
concentration in the two samples is approximately equal so that the results suggest that while excess nitrogen reduces
considerably the intensity of the PL emission, presumably by forming other complexes with Ni, some of the Ni
nevertheless enters the host lattice as $1.4$~eV Ni color centers.

\begin{figure}[t]
\begin{center}
\includegraphics[width=1\linewidth]{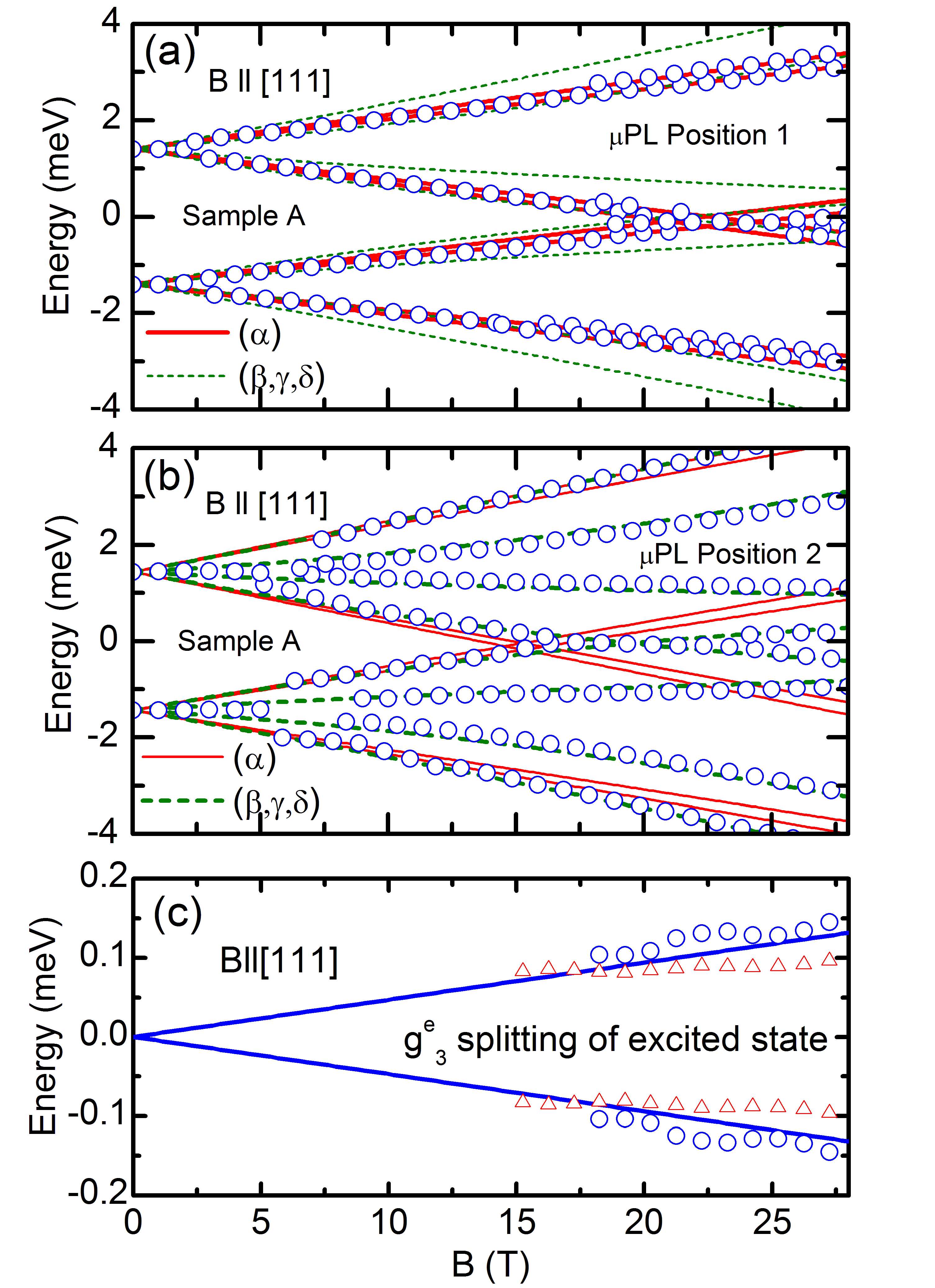}
\end{center}
\caption{(Color on-line) (a-b) Energy of the observed transitions (symbols) for $T=4.2$~K $\mu$-photoluminescence
collected at two different positions on the $(111)$ face with $B\parallel[111]$ on sample A as a function of the
magnetic field. The predictions of the NND effective spin Hamiltonian are shown for a defect preferentially aligned
along $[111]$ (red solid lines) and the other three equivalent $\langle111\rangle$ directions (green broken lines).
Note that at Position 2 the defects are not preferentially aligned along the growth direction and fitting requires a
radically different set of parameters. (c) Splitting of the excited state; the high (circles) and low (triangles)
energy transitions in (a), from which the average energy of the transition has been subtracted, are plotted versus
magnetic field. The solid lines are a least squares fit to extract $g^e_3=0.16\pm0.03$.}\label{fig10_micro111}
\end{figure}

Fig.~\ref{fig10_micro111}(a-b) shows $\mu$-PL measurements obtained at two different positions on the (111) face of
sample A with $B\parallel[111]$. We have performed a full map of the face and position~$1$ in
Fig.~\ref{fig10_micro111}(a) is representative of most of the (111) face. The behavior is identical to that observed in
macro-PL. The predictions of the effective spin Hamiltonian for a defect preferentially aligned along the $[111]$
direction are shown by the red solid lines and the other orientations by the green broken lines. The data is well
fitted by the preferentially aligned defect scenario. As emission from position~$1$ is representative of the (111) face
this allows us to conclude that the defect is preferentially aligned along $[111]$ growth direction over most of the
(111) face. The splitting of the excited state is also clearly seen in this high resolution low noise data taken in
static magnetic fields. See \emph{e.g.} the splitting of the lowest and highest energy transition in
Fig.~\ref{fig10_micro111}(a) above $15$~T. Fitting the Hamiltonian to this splitting it is possible to extract a
refined value of $g_3^e$ which describes the small spin splitting of the excited state with the field applied along the
$Z$ quantization axis. In Fig.~\ref{fig10_micro111}(c) the excited state splitting is shown; we plot the energy of the
split transitions after subtracting the average energy to remove the background. A least squares fit to both data sets
(solid lines) gives $g_3^e=0.16\pm0.03$. The complete set of parameters for the effective spin Hamiltonian is given in
Table~\ref{tab1}. The values proposed by NND are shown for comparison. The differences are small showing both the
reproducibility of the results between different samples and the remarkable job performed by NND from limited PL data
in relatively low magnetic fields $B \leq 6$~T.

\begin{figure}[t]
\begin{center}
\includegraphics[width=1\linewidth]{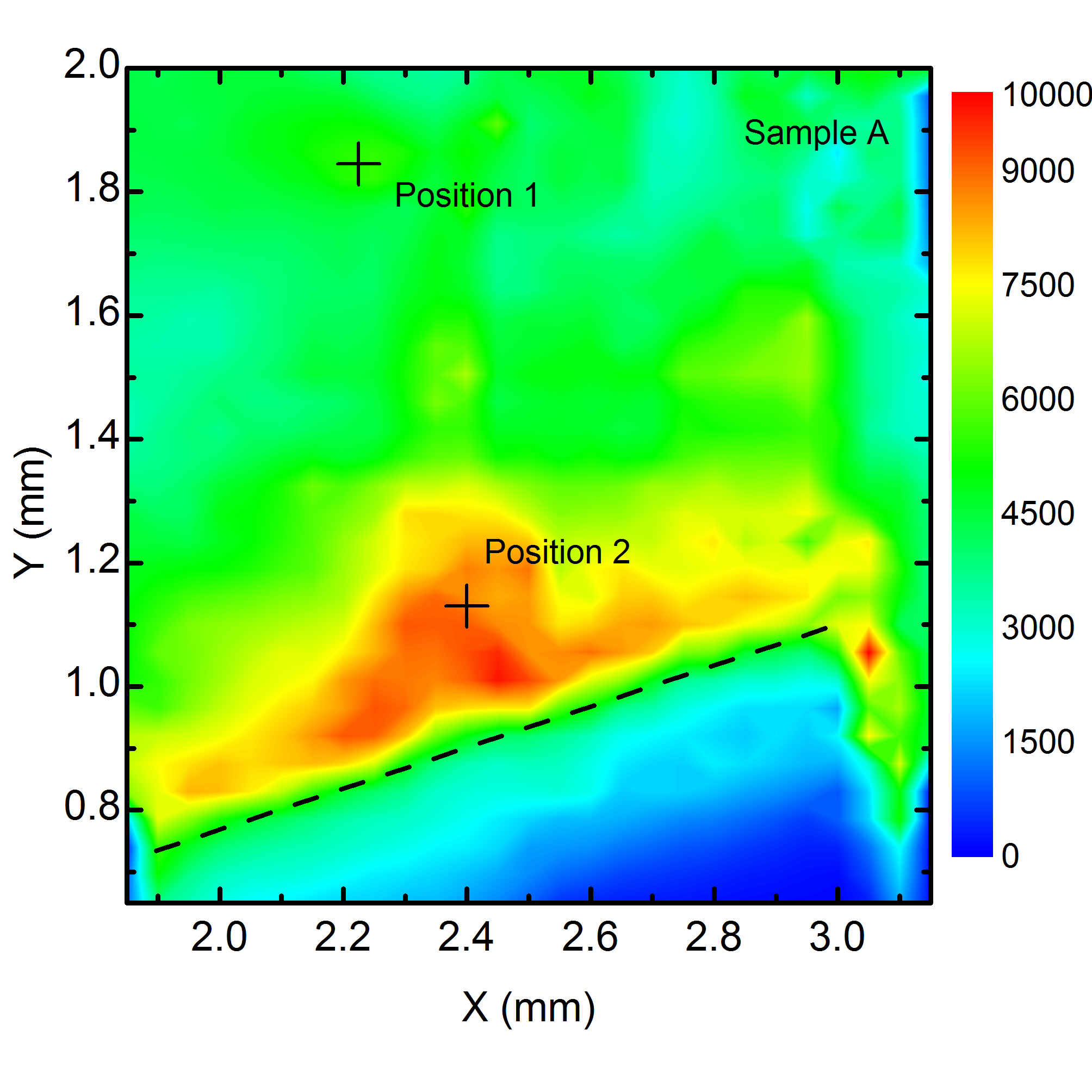}
\end{center}
\caption{(Color on-line) Spatial PL map of the $(111)$ face measured at $T=4.2$~K and $B=0$~T showing the integrated
intensity of the high energy component of the $1.4$~eV doublet. The dashed line indicates the bottom of the (111) face.
Intensity decreases rapidly below this line due to the loss of focus. The positions at which the $\mu$-PL spectra in
Fig.~\ref{fig5_PL_0T} were taken are indicated.}\label{fig11_111map}
\end{figure}

It is however possible, towards the bottom of the $(111)$ face \emph{i.e} when going far away from the $[1\bar10]$ top
edge (see Fig.\ref{fig4_diamond}), to find very different spectra as shown in Fig.\ref{fig10_micro111}(b).
\begin{table*}[]
\begin{center}
\begin{tabular}{c | c c c c c c}
 & $g_3$ & $g_3^{'}$ & $g_1$ & $g_3^e$ & $g_1^e$ & $\lambda$~(meV)\\
\hline
Isoya ESR Ref.[\onlinecite{Isoya90}] & 2.329 & -- & -- & -- & -- & --\\
NND PL Ref.[\onlinecite{Nazare91}] & 2.42 & 1.62 & 1.28 & 0.18 & 2.5 & 2.8\\
Mason MCDA Ref.[\onlinecite{Mason99}] & 2.32 & -- & -- & $<0.1$ & 2.445 & --\\
Maes PL Ref.[\onlinecite{Maes04}] & (2.329) & 1.93 & -- & -- & -- & --\\
Standard parameters (this work) & $2.3 \pm 0.05$ & $2.0 \pm 0.05$ & $1.7 \pm 0.05$ & $0.16 \pm 0.03$ & $2.4 \pm 0.05$ & 2.800\\
Defect misaligned on $(111)$ face & $3.5 \pm 0.1$ & $3.0 \pm 0.1$ & $1.7 \pm 0.1$ & (0.16) & $2.3 \pm 0.1$ & 2.874\\
\end{tabular}
\end{center}
\caption{Parameters of the effective $S=1/2$ spin Hamiltonian found here and in previous work. The standard parameters
fit the majority of the present data. The misaligned parameter set corresponds to the special case of defects on the
$(111)$ face which do not have their quantization axis preferentially aligned along the $[111]$ growth directions. The
larger estimated error for the misaligned defects is due to the restricted data set which is limited to the
$B\parallel[111]$ configuration on the $(111)$ face. Note, the NND values are twice those given in
Ref.[\onlinecite{Nazare91}] due to their implicit use of an $S=1$ Hamiltonian when writing the secular
equation.}\label{tab1}
\end{table*}
The spectra measured at position~$2$ contain many more lines indicating that the defect is not preferentially aligned
along the growth direction. Such spectra occur in areas of the $(111)$ with increased intensity of emission in zero
magnetic field (see Fig.\ref{fig11_111map}). In addition a radically different parameter set for the effective spin
Hamiltonian is required to fit the data. We have fitted the data assuming that the defect orientation is not along the
$[111]$ growth direction (green broken lines). For $B\parallel[111]$ the three ``misaligned'' defect orientations are
all equivalent. The extracted parameters are given in Table~\ref{tab1}. The fit is almost perfect confirming the
trigonal symmetry. The red solid lines are the prediction for a defect aligned along the $[111]$ growth direction. From
the data above $20$~T it appears that these transitions are completely absent from the spectra. This suggests that the
defect may even be preferentially ``not aligned'' with the growth direction in this region of the face which is close
to the seed crystal. Moreover, the parameters required to fit are considerably different from those obtained
previously. In particular, the values of $g_3$ and $g_3^{'}$ are $\sim 50$\% larger reflecting the much larger Zeeman
splitting of the ground state even though the magnetic field is not aligned along the $Z$ quantization axis. This
suggests that the local environment (trigonal distortion) of the Ni center is significantly different in regions of the
$(111)$ face where it does not preferentially align along the $[111]$ growth direction.

\section{Discussion}

\begin{figure}[t]
\begin{center}
\includegraphics[width=1\linewidth]{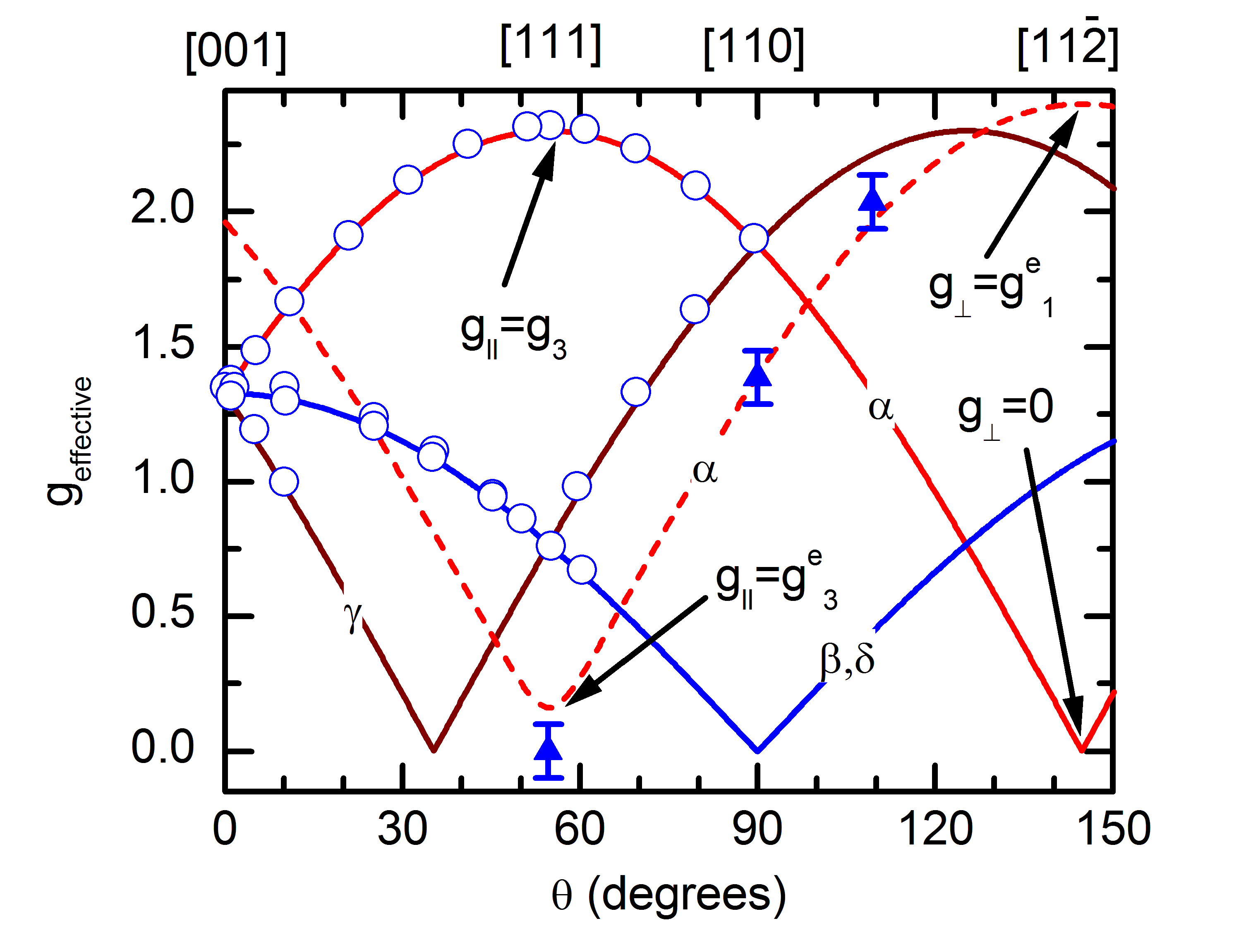}
\end{center}
\caption{(Color on-line) The circles show the angular dependence of the line position (effective $g$-factor) of the
NIRIM-2 center, taken from Ref.[\onlinecite{Isoya90}], with the magnetic field rotated about the $[\bar110]$ axis at
$T=4$~K and $B\simeq0.5$~T. The triangles are the splitting of the excited state measured using MCDA taken from
Ref.[\onlinecite{Mason99}]. Also shown is the predicted angular dependence of the ground state splitting (solid lines)
and excited state splitting (broken lines), calculated using the effective spin Hamiltonian with the parameters found
in this work, and given in Table~\ref{tab1}. The curves are labeled according to the alignment of the defect $Z$
quantization axis.}\label{fig12_Isoya}
\end{figure}

The $1.4$~eV Ni center investigated here has been unambiguously identified with the NIRIM-2 ESR line in HPHT
diamond.~\cite{Davies89,Nazare91} NIRIM-2 has a large angular dependence of the magnetic field position of the ESR
lines which is consistent with the center having trigonal symmetry.\cite{Isoya90} ESR gives very precise values for the
g-factors but probes only the splitting of the ground state. While magnetic circular dichroism (MCDA) is a less precise
technique, it is extremely useful since it gives access to the $g$-factors of the excited state.~\cite{Mason99} The
measured anisotropy of the $g$-factor can provide a crucial test for a given microscopic model. In
Fig.~\ref{fig12_Isoya} we show the ground state splitting (effective $g$-factor) obtained from the ESR data of Isoya
\emph{et al.}~\cite{Isoya90} together with the MCDA results of Mason \emph{et al.}~\cite{Mason99} for the excited state
splitting. The expected angular dependence of the ground state splitting using the effective spin Hamiltonian is also
plotted for the ground (solid lines) and excited state (broken line) as a function of angle when the magnetic field is
rotated around the $[\bar110]$ direction, \emph{e.g.} from $B\parallel[001]$ to $B\parallel[11\bar2]$. The predictions
of the effective spin Hamiltonian are in excellent agreement with the ESR and MCDA results for both the ground and
excited states.

In order to compare the $g$-factors of the effective spin Hamiltonian with the ESR and MCDA results, it is convenient
to define effective $g$-factors, $g_\parallel$ and $g_\perp$ corresponding to a magnetic field aligned parallel and
perpendicular to the $Z$ quantization axis of the defect. For the ground state we have a large $g_\parallel=g_3$ and
$g_\perp=0$. The magnetic field component which is not along $Z$ does not generate a Zeeman splitting of the ground
state. The $B_X$ and $B_Y$ components only change the splitting of the zero field doublet as can be seen from
Eq.~\ref{Eq_eigenvalues}. The situation for the excited state is very different with a small $g_\parallel=g^e_3$ and a
large $g_\perp = g^e_1$. As pointed out by Mason \emph{et al.}, the very small value of $g_\parallel\approx0$ is a most
unexpected result for an orbital singlet state.\cite{Mason99} The high magnetic field PL results presented here have
allowed a rather precise determination of the $g$-factors which motivates us to revisit Mason \emph{et al.}'s
quantitative comparison with the $g$-factors derived from the crystal field theory of interstitial 3d$^9$ Ni$^+$.

\subsection{Crystal field model}

\begin{figure}[]
\begin{center}
\includegraphics[width=0.8\linewidth]{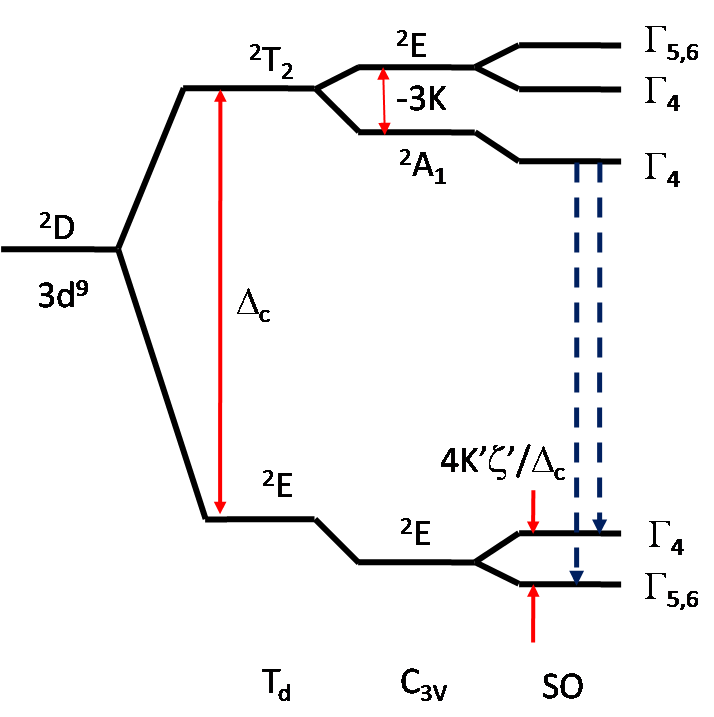}
\end{center}
\caption{(Color on-line) Energy levels of the $^2D$ state of $3d^9$ interstitial Ni$^+$ showing the subsequent effects
of the cubic ($T_d$) and trigonal ($C_{3v}$) crystals fields and spin-orbit interaction.~\cite{Mason99} The broken
arrows indicate the observed zero field optical transitions.}\label{fig13_levels}
\end{figure}

It has been proposed that the Hamiltonian used for neutral substitutional vanadium in SiC\cite{Kaufmann97} can also be
applied to the case of interstitial Ni$^+$,\cite{Mason99}
\begin{equation}
H=H_{T_d} + H_{C_{3v}} + H_{S.O} + H_Z,
\end{equation}
where $H_{T_d}$ contains the cubic crystal field term, $H_{C_{3v}}$ the trigonal crystal field, $H_{S.O}$ the
spin-orbit interaction and finally $H_Z$ includes the Zeeman interaction. The resulting level structure is shown in
Fig.~\ref{fig13_levels}. The parameters $\Delta_c$ , $K$, $\zeta$, and $k$ are the cubic crystal-field splitting,
one-third of the trigonal crystal-field splitting within the $^2T_2$ state, the spin-orbit parameter, and an orbital
reduction factor, respectively ($K'$, $\zeta'$, and $k'$ are the corresponding quantities for the $^2E$ state). The
sign of the $^2T_2$ trigonal crystal field parameter, $K<0$, is chosen so that the $A_1$ level of the excited state is
lowest in agreement with the uniaxial stress measurements~\cite{Nazare91} and the spin-orbit parameters $\zeta$ and
$\zeta'$ are both negative for the single $d^9$ hole.

The $^2E$ ground state is split into two Kramers doublets by a combination of spin-orbit coupling and trigonal
distortion.
\begin{equation}\label{Eq_spinorbit}
\lambda = E(\Gamma_4)-E(\Gamma_{5,6})=\frac{4 \zeta' K'}{\Delta_c}
\end{equation}
where the cubic crystal field term $\Delta_c\simeq1.4$~eV. From the measured splitting $\lambda=2.8$~meV we obtain the
product $\zeta' K'=0.98 \times 10^{-3}$~eV$^2$.

The $g$-factors of the ground state doublet are for the $\Gamma_{5,6}$ state,~\cite{Mason99}
\begin{multline}\label{Eq_g_ground1}
g_{\parallel} = g_3 = 2 - \frac{4k'\zeta'}{\Delta_c} - \frac{8k'K'}{\Delta_c},\\
g_\perp=0,\\
\end{multline}
and for the $\Gamma_{4}$ state
\begin{multline}\label{Eq_g_ground2}
 g_{\parallel} = g'_3 = 2 - \frac{4k'\zeta'}{\Delta_c} + \frac{8k'K'}{\Delta_c},\\
g_\perp=\frac{4k'\zeta'}{\Delta_c}.\\
\end{multline}
As $\Delta_c \gg K', \zeta'$ the  $\Gamma_{4}$ state has $g_\perp \approx 0$ so that the effective spin Hamiltonian (in
which \emph{de facto} $g_\perp=0$) is a reasonable approximation. Using the values of $g_3=2.3$ and $g'_3=2.0$ found in
this work, solving Eqs.(\ref{Eq_g_ground1}-\ref{Eq_g_ground2}) immediately gives
$4k'\zeta'/\Delta_c=8k'K'/\Delta_c=-0.15$ so that $g_\perp = -0.15$. Although small, the splitting of the upper doublet
state should be experimentally observable at magnetic fields above $15$~T with $B$ applied perpendicular to the $Z$
defect axis on the $[111]$ face. The high resolution $\mu$-PL measurements, with its narrower line widths and better
signal/noise in static magnetic fields, clearly resolved the similarly small splitting of the excited state with $B
\parallel Z$. Unfortunately, only the Faraday configuration is possible when using the $\mu$-PL system so that such a
measurement is excluded for the present. Making the approximation that the orbital reduction factor $k'\approx1$ gives
$K'=\zeta'/2=0.0262$~eV. As pointed out by Mason \emph{et al.}, while such values of  $\zeta'$ and $K'$ are reasonable,
the product $\zeta' K'=1.4 \times 10^{-3}$~eV$^2$ is too large to be compatible with the $\zeta' K'\simeq1 \times
10^{-3}$~eV$^2$ obtained from Eq.(\ref{Eq_spinorbit}) knowing the spin-orbit splitting $\lambda=2.8$~meV of the zero
field doublet. Using an orbital reduction factor $k'<1$ only makes the situation worse.

The $g$-factors of the excited state (lower $\Gamma_4$ of $^2T_2$) are given by,~\cite{Mason99}
\begin{multline}\label{Eq_g_excited}
g_{\parallel}= g^e_3 = 2(a^2-b^2)-2kb^2\\
g_\perp = g^e_1 = 2a^2-2\sqrt{2}kab,\\
\end{multline}
where,
\begin{multline}\label{Eq_cossintan}
a=\cos(\gamma),\\
b=-\sin(\gamma),\\
\gamma=\frac{\zeta/\sqrt{2}}{3K/2 - \zeta/4}.\\
\end{multline}

The $\cos$ and $\sin$ terms give an additional constraint that $a^2 + b^2 =1$, so that the orbital reduction factor $k$
cannot be freely chosen. Using the values of $g^e_3=0.16$ and $g^e_1=2.4$ found in this work, a solution to
Eq.(\ref{Eq_g_excited}) exists with $k=0.79$, $a=0.82$ and $b=-0.574$. From Eq.(\ref{Eq_cossintan}) this implies that
$K \simeq \zeta/3.5$. Neglecting covalency effects, which are expected to be small in diamond, $\lambda\approx\lambda'$
giving $K \approx 0.015$~eV.

Thus the only apparent short coming of the crystal field model is its inability to correctly predict the zero magnetic
field doublet splitting. It has been suggested that this might be explained by corrections to the splitting of the
$^2E$ state which can arise due to a mixing with the $4p$ states of the Ni$^+$.~\cite{Paslovsky92} Such a mechanism has
also been proposed for $3d^9$ Cu$^{2+}$ in II-VI compounds.~\cite{Scherz69,Telahun96}

A crucial test of the crystal field model is provided by defects with a quantization axis which is not preferentially
aligned: In the $\mu$-PL data it is possible to find regions of the $(111)$ face, towards the bottom of the sample,
\emph{i.e.} near the seed crystal, where the defects are not preferentially aligned along the $[111]$ growth direction.
Such defects have an almost identical emission energy (see Fig.~\ref{fig5_PL_0T}) and an almost identical splitting
($\lambda=2.874$~meV) of the zero magnetic field doublet. Surprisingly, the $g$-factors for the trigonal axis of such
defects are markedly different (see Table.~\ref{tab1}). Using the values found in experiment, $g_3=3.5$ and $g'_3=3.0$
in Eqs.(\ref{Eq_g_ground1}-\ref{Eq_g_ground2}), gives $4k'\zeta'/\Delta_c=-1.25$ and $8k'K'/\Delta_c=-0.25$ so that
$K'=\zeta'/10$. Assuming $k'\approx0$, the spin-orbit coupling term $\zeta' \simeq 0.44$~eV and trigonal crystal field
term $K' \simeq 0.044$~eV. The value of $\zeta'$ is an unreasonable $6$ times larger that the value of $0.075$~eV for a
free Ni$^+$ ion.\cite{Mason99,Moore49} Moreover, the product $\zeta' K'\simeq 20 \times 10^{-3}$~eV$^2$ overestimates
the zero field doublet splitting (Eq.(\ref{Eq_spinorbit})) by over an order of magnitude. At the same time experiment
indicates that the splitting of the zero field doublet is almost unchanged. This forces us to conclude that the crystal
field model does not provide a correct quantitative description of the $1.4$~eV Ni color center in diamond.

\section{Conclusion}

We have investigated two different boron-free HTHP synthetic diamond crystals. Our high magnetic field
photoluminescence results are perfectly described by the NND effective spin Hamiltonian, unequivocally demonstrating
the trigonal symmetry of the $1.4$~eV Ni color center in diamond. Both samples investigated have a similar Ni content
($\simeq 1\times 10^{19}$ cm$^{-3}$) but radically different concentrations of N. Sample B has a similar concentrations
of Ni and N, while sample A has roughly two orders of magnitude less nitrogen. Magnetization measurements show that Ni
is predominantly incorporated into Sample B as isolated Ni complexes and into Sample A as nm size Ni clusters.
Nevertheless, both samples exhibit the characteristic $1.4$~eV emission doublet associated with an isolated Ni complex.
Despite the significantly lower concentration of isolated Ni, Sample A shows much stronger $1.4$~eV emission. This
suggests that the presence of N does not necessarily favor the formation of the $1.4$~eV Ni color center; the
concentration of isolated Ni-N complexes is almost certainly higher in Sample B, while the $1.4$~eV PL signal is much
weaker, suggesting that the presence of N may actually impede the formation of this particular Ni color center. As
nitrogen is usually a donor in diamond, a possible mechanism could be the transfer of a donor electron to the Ni$^+$
ion reducing the number of optically active centers.\cite{Paslovsky92,Collins90}

The magneto-PL presented here, together with previously published magneto-PL and PL under uniaxial
stress,\cite{Nazare91,Maes04} ESR~\cite{Isoya90} and MCDA\cite{Mason99} are all consistent with an interstitial $3d^9$
Ni$^+$ with spin $S=1/2$ and a large trigonal distortion ($C_{3v}$ symmetry) due to a displacement of the Ni along a
$<111>$ direction. The exact nature of the complex nevertheless remains unknown. A complex involving an additional,
transition metal impurity, dopants such as nitrogen or boron, or a vacancy or divacancy, is required to produce the
trigonal distortion. On the basis of the ESR results Isoya \emph{et al.} suggested an interstitial $Ni{^+}$-vacancy
complex. However, this configuration has been shown to be unstable~\cite{Larico04} with the interstitial Ni moving
towards vacancy \emph{i.e.} to the substitutional site. First principles calculations suggests that that a complex
involving boron and substitutional nickel fulfills all the necessary requirements.\cite{Larico09} However, this
assignment seems to be unlikely here as there was no boron present in the melt during the growth of our samples.
Moreover, recent density functional theory (DFT) calculations coupled with X-ray absorption spectroscopy measurements
performed on Sample B, suggest the incorporation of nickel as a divacancy complex, in which interstitial Ni is placed
at the midpoint between two vacancies.\cite{Gheeraert12} However, the first principles calculations of Larico \emph{et
al.}~\cite{Larico09} suggest that the (VNiV)$^+$ complex has $C_{2h}$ symmetry with the Ni related electronic states
resonant and inert inside the valence band. First principles calculations of defects in diamond are in general
complicated due to the need to take into account possible relaxations of neighboring carbon atoms which can alter both
the energy and the symmetry of the center. Under such conditions the microscopic model for the $1.4$~eV Ni color center
in diamond should still be considered as an open question.

\begin{acknowledgments}
The help of Fabrice Donatini with the cathodoluminescence measurements is gratefully acknowledged. We thank Laurence
Eaves and Bernard Clerjaud for their interest in this work and stimulating discussions. This work was partially
supported by EuroMagNET II under the EU contract 228043.
\end{acknowledgments}

%\bibliography{diamond}


\end{document}